\documentclass[fleqn,twoside]{article}
\usepackage{espcrc2}

\usepackage{graphicx}
\usepackage[figuresright]{rotating}

\def\ie{{\it i.e.}}
\def\eg{{\it e.g.}}

\def\to{\rightarrow}
\def\sub#1{_{\lower.25ex\hbox{$\scriptstyle#1$}}}
\def\tev{\,{\ifmmode\mathrm {TeV}\else TeV\fi}}
\def\gev{\,{\ifmmode\mathrm {GeV}\else GeV\fi}}
\def\mev{\,{\ifmmode\mathrm {MeV}\else MeV\fi}}
\def\mpl{\ifmmode M_{pl}\else $M_{pl}$\fi}
\def\mpl{\ifmmode \overline M_{Pl}\else $\bar M_{Pl}$\fi}
\def\to{\rightarrow}

\def\subw{_{\rm w}}
\def\mh{\ifmmode m\sbl H \else $m\sbl H$\fi}
\def\mch{\ifmmode m_{H^\pm} \else $m_{H^\pm}$\fi}
\def\mt{\ifmmode m_t\else $m_t$\fi}
\def\mc{\ifmmode m_c\else $m_c$\fi}
\def\mz{\ifmmode M_Z\else $M_Z$\fi}
\def\mw{\ifmmode M_W\else $M_W$\fi}
\def\mws{\ifmmode M_W^2 \else $M_W^2$\fi}
\def\mhs{\ifmmode m_H^2 \else $m_H^2$\fi}
\def\mzs{\ifmmode M_Z^2 \else $M_Z^2$\fi}
\def\mts{\ifmmode m_t^2 \else $m_t^2$\fi}
\def\mcs{\ifmmode m_c^2 \else $m_c^2$\fi}
\def\mchs{\ifmmode m_{H^\pm}^2 \else $m_{H^\pm}^2$\fi}
\def\ztwo{\ifmmode Z_2\else $Z_2$\fi}
\def\zone{\ifmmode Z_1\else $Z_1$\fi}
\def\mtwo{\ifmmode M_2\else $M_2$\fi}
\def\mone{\ifmmode M_1\else $M_1$\fi}
\def\tb{\ifmmode \tan\beta \else $\tan\beta$\fi}
\def\xw{\ifmmode x\subw\else $x\subw$\fi}
\def\ch{\ifmmode H^\pm \else $H^\pm$\fi}
\def\lum{\ifmmode {\cal L}\else ${\cal L}$\fi}
\def\inpb{\,{\ifmmode {\mathrm {pb}}^{-1}\else ${\mathrm {pb}}^{-1}$\fi}}
\def\infb{\,{\ifmmode {\mathrm {fb}}^{-1}\else ${\mathrm {fb}}^{-1}$\fi}}
\def\epem{\ifmmode e^+e^-\else $e^+e^-$\fi}
\def\ppb{\ifmmode \bar pp\else $\bar pp$\fi}
\def\bsg{\ifmmode B\to X_s\gamma\else $B\to X_s\gamma$\fi}
\def\bsll{\ifmmode B\to X_s\ell^+\ell^-\else $B\to X_s\ell^+\ell^-$\fi}
\def\bstt{\ifmmode B\to X_s\tau^+\tau^-\else $B\to X_s\tau^+\tau^-$\fi}
\def\lamt{\ifmmode \tilde\lambda\else $\tilde\lambda$\fi}
\def\shat{\ifmmode \hat s\else $\hat s$\fi}
\def\that{\ifmmode \hat t\else $\hat t$\fi}
\def\uhat{\ifmmode \hat u\else $\hat u$\fi}

\newskip\zatskip \zatskip=0pt plus0pt minus0pt
\def\matth{\mathsurround=0pt}
\def\lsim{\mathrel{\mathpalette\atversim<}}
\def\gsim{\mathrel{\mathpalette\atversim>}}
\def\atversim#1#2{\lower0.7ex\vbox{\baselineskip\zatskip\lineskip\zatskip
  \lineskiplimit 0pt\ialign{$\matth#1\hfil##\hfil$\crcr#2\crcr\sim\crcr}}}

\def\grtsim{\,\,\rlap{\raise 3pt\hbox{$>$}}{\lower 3pt\hbox{$\sim$}}\,\,}
\def\lsim{\,\,\rlap{\raise 3pt\hbox{$<$}}{\lower 3pt\hbox{$\sim$}}\,\,}

\newcommand{\AmS}{{\protect\the\textfont2
  A\kern-.1667em\lower.5ex\hbox{M}\kern-.125emS}}

\title{No Prejudice in Space}

\author{R.C.~Cotta, J.S.~Gainer, J.L.~Hewett and T.G.~Rizzo\address{SLAC 
        National Accelerator Laboratory, \\ 
        2575 Sand Hill Rd, Menlo Park, CA, 94025, USA}%
\thanks{Presented at the {\it Dark Matter Conference},
9-11 Feb 2009, Arcetri, Florence, Italy.
SLAC-PUB-13731.  Work supported in part by the Department of Energy, 
                 Contract DE-AC02-76SF00515.}}
        
\begin{document}
\vspace{-1.5cm}

\begin{abstract}
We present a summary of recent results obtained 
from a scan of the 19-dimensional parameter space of the pMSSM and its 
implications for dark matter searches.

\vspace{1pc}
\end{abstract}

\maketitle

\section{Introduction} 

Supersymmetry (SUSY) is a leading candidate for a theory of physics
beyond the Standard Model.
However, it is clear that if SUSY exists, it must be broken. 
The mechanism by which SUSY is broken is yet unknown 
and there is a growing list of possible scenarios.   
In these scenarios, the SUSY spectrum is described by a
handful of parameters, generally defined at the SUSY breaking scale;
RGE running of sparticle masses and coupling constants
yields predictions for the mass spectra and decay patterns of the
various sparticles at energy scales relevant for colliders or
cosmology.  However, these SUSY breaking scenarios are restrictive and
predict specific phenomenologies for colliders and cosmology that
may not represent the full range of possible SUSY signatures.

Here, we study the MSSM more broadly without
assumptions at the high scale. We restrict ourselves
to the CP-conserving MSSM (\ie, no new phases) with minimal flavor
violation (MFV)\cite{mfv}.  Additionally, we require that the first two
generations of sfermions be degenerate as motivated by constraints
from flavor physics. We are then left with 19  independent, real,
weak-scale, SUSY Lagrangian parameters: the gaugino masses
$M_{1,2,3}$, the Higgsino mixing parameter $\mu$, the ratio of the
Higgs vevs $\tan \beta$, the mass of the pseudoscalar Higgs boson
$m_A$, and the 10 squared masses of the sfermions ($m_{\tilde{q}1,3}$,
$m_{\tilde{u}1,3}$,$m_{\tilde{d}1,3}$,$m_{\tilde{l}1,3}$,and
$m_{\tilde{e}1,3}$).  We include independent $A$-terms only for
the third generation ($A_b$, $A_t$, and $A_\tau$) due to the small
Yukawa couplings for the first two generations.  This set of 19
parameters has been called the phenomenological MSSM
(pMSSM){\cite{Djouadi:2002ze}}.

To study the pMSSM, we performed a scan over this 19-dimensional
parameter space assuming flat priors for the specified 
ranges\cite{Berger:2008cq}:  
$100 \gev \leq m_{\tilde f} \leq 1\tev;~50\gev \leq |M_{1,2},\mu|
\leq 1 \tev;~100 \gev \leq M_3\leq 1 \tev;~|A_{b,t,\tau}| \leq 1 
\tev;~1 \leq \tan \beta \leq 50;~43.5\gev \leq m_A \leq 1 \tev.$
We randomly generated $10^7$ points in this parameter space and subjected 
them to an exhaustive set of
existing theoretical and experimental constraints.  We
also performed a similar scan with log priors (with slightly different mass
ranges) to gauge the influence of priors on our results and found these 
to be negiglible\cite{Berger:2008cq}.
We then generated SUSY spectra utilizing
SuSpect2.34\cite {Djouadi:2002ze}.

\section{Theoretical and Experimental Constraints}

We now discuss the theoretical and experimental constraints that we
applied to the generated parameter space points; for more details, one should
consult~\cite{Berger:2008cq,us2}.

\subsection{Theoretical Constraints}

We demanded that the sparticle spectrum not have tachyons or color or
charge breaking minima in the scalar potential and also required  
that the Higgs potential be bounded from below with consistent 
electroweak symmetry breaking. 
We assume that the LSP, which will be absolutely stable, be a
conventional thermal relic and identify the LSP as the lightest neutralino.

\subsection{Low Energy Constraints}

The code  micrOMEGAs2.20{\cite {MICROMEGAS}} was used to evaluate the
following observables for each point in the parameter space:
$\Delta \rho$, the decay rates for $b\to s\gamma$ and $B_s \to
\mu^+\mu^-$, and the $g-2$ of the muon.  In addition, we evaluated the
branching fraction for $B\to \tau \nu$ following{\cite{gino}}
and {\cite {ems}}.  
We allowed a large range for the SUSY
contribution to $g-2$ due to the evolving discrepancy
between theory and experiment{\cite{Bennett:2006fi}}.
We implemented constraints from meson-antimeson mixing{\cite{mesonmix}}
by assuming MFV{\cite{mfv}}, imposing first and second generation mass
degeneracy, and  demanding that the ratio of first/second to
third generation squark soft breaking masses 
differ from unity by no more than a factor of $5$.

\subsection{Accelerator Constraints}

LEP data at the $Z$ pole shows that 
charged sparticles with masses below $M_Z/2$ are unlikely. This also 
holds for the lightest neutral Higgs boson.  Data from 
LEPII{\cite {lepstable}} indicates that there are no new {\it stable} charged
particles of any kind with masses below 100 GeV.   We also require
that any new contributions to the
invisible width of the $Z$ boson be $\leq 2$ MeV{\cite {LEPEWWG}}.

Results from sparticle searches at LEPII possess numerous caveats.
We implement a lower limit of $92$ GeV
on first and second generation squark masses{\cite {ALEPH}} and 95 GeV
on the sbottom mass\cite{bbb},
provided that the gluino
is more massive than the squarks and the mass difference ($\Delta m$)
between the squark and the LSP is $\geq 10$ GeV.  We
demand that the lightest stop mass be greater than $95(97)$ GeV\cite{LEPSUSY}
if the stop can(cannot) decay into  $Wb\chi_1^0$.
The right-handed sleptons must have
masses greater than $100$, $95$, or $90$ GeV for selectrons, smuons,
and staus respectively, as long as the 
condition $0.97 m_{\mathrm{slepton}} > m_{\mathrm{LSP}}$ is satisfied.  
These bounds are also applied to
left-handed sleptons, when the neutralino $t-$channel diagram
may be neglected in the case of selectrons.
Chargino masses be greater than $103(95)$ GeV, provided
that the LSP-chargino mass splitting is $\Delta
m > (<) 2$ GeV\cite{LEPSUSY}.  If the
lightest chargino is dominantly Wino, this limit only applies
when the electron sneutrino $t-$channel diagram is negligible.
The LEP Higgs Working Group{\cite {LEPHIGGS}},
provides five sets of constraints on the MSSM Higgs sector, which are
essentially limits on the Higgs-Z coupling times
the Higgs branching fraction for various final states.
We employ SUSY-HIT\cite{SUSYHIT} to analyze these.  In addition,
we included a theoretical uncertainty on the calculated mass of the
lightest Higgs
boson of approximately $3$ GeV{\cite {uncertain}} when applying these
constraints.

We also employ constraints from searches at the Tevatron. 
Restrictions on the squark and gluino sectors arise from
the null D0 multijet plus missing energy search{\cite{domet}}. 
We generalize their analysis, rendering it model independent, by
generating multijet plus missing energy events for our models using
PYTHIA6.4{\cite {PYTHIA}} as interfaced to PGS4
{\cite  {PGS}} which provides a fast detector simulation.
We weigh our results with K factors computed using PROSPINO2.0
{\cite  {PROSPINO}}. 
Analogously, we employ constraints from the CDF search for trileptons
plus missing energy{\cite {cdftrilepton}}, which we also generalize to
the full pMSSM. 
D0{\cite {dostable}} has obtained lower limits on the mass of
heavy stable charged particles.  
We take this constraint to be $m_{\chi^+}\geq 206 |U_{1w}|^2 +171
|U_{1h}|^2$ GeV at $95\%$ CL for charginos, 
where the matrix entries $U_{1w}$ and
$U_{1h}$ determine the Wino/Higgsino content of the lightest chargino. 
CDF and D0 also have analyses that search for light stops and
sbottoms{\cite {stops}}; these searches are difficult to implement
in a model-independent pMSSM context and thus we exclude models with
light ($m < m_t$) stops or sbottoms. 
  
\subsection{Astrophysical Constraints}

There are two constraints from considering the LSP as a long-lived
relic.  As noted above, we demand that the LSP be the lightest
neutralino.  We also require, following the 5 year WMAP
measurement{\cite {Komatsu:2008hk}} of the relic density, that $\Omega
h^2|_{\mathrm{LSP}} \leq 0.121$.
In not employing a lower bound on $\Omega h^2|_{\mathrm{LSP}}$ for our
models, we acknowledge the possibility
that even within the MSSM and the thermal relic framework, dark matter
may have multiple components. 
However, in discussing results below, we will also examine a subset of
models for which $0.100 \leq \Omega h^2|_{LSP} \leq 0.121$.

We also obtain constraints from direct dark matter
searches{\cite{dmsearch}}.
Generally, the strongest constraints come from the spin-independent
WIMP-nucleon cross sections, hence we only implement bounds on our
models from these; inspection of the spin-dependent WIMP-nucleon cross
sections in our models confirms that this approach is reasonable.
Both spin-independent and spin-dependent cross sections were
calculated using micrOMEGAs2.21{\cite {MICROMEGAS}}.
We implement cross section limits from XENON10{\cite
  {XENON10}}, CDMS{\cite {CDMS}}, CRESST I{\cite {CRESST}} 
and DAMA{\cite {DAMA}} data. It
should be noted that many of our models predict a value
$\Omega h^2|_{\mathrm{LSP}}$ which is less than that observed by WMAP
and supernova searches.  We thus scale our cross sections 
to take this into account.

\section{Results}

As noted above, we randomly generated $10^7$ parameter space points (\ie,
models) in a 19-dimensional pMSSM parameter space using flat priors.
Only $\sim 68.5 \cdot 10^3$ of these models satisfy all the
constraints listed above.  The properties of these
models are described in much greater detail in~\cite{Berger:2008cq}.  Here
we will discuss the attributes of these models
which are most important astrophysically.

\begin{figure}[t]
\begin{center}
\includegraphics[width=8.5cm,angle=0]{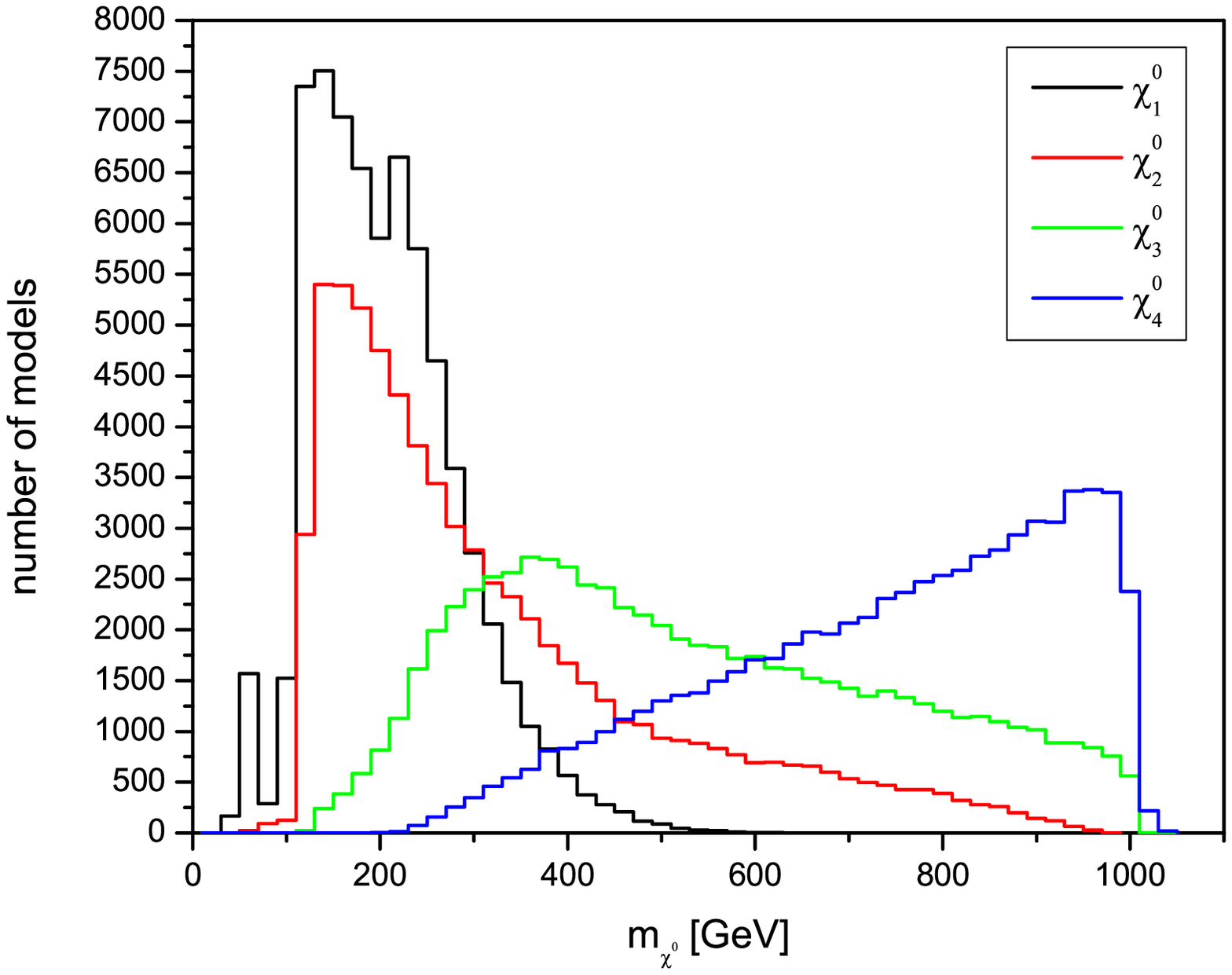}
\vspace*{-0.1cm}
\includegraphics[width=8.5cm,angle=0]{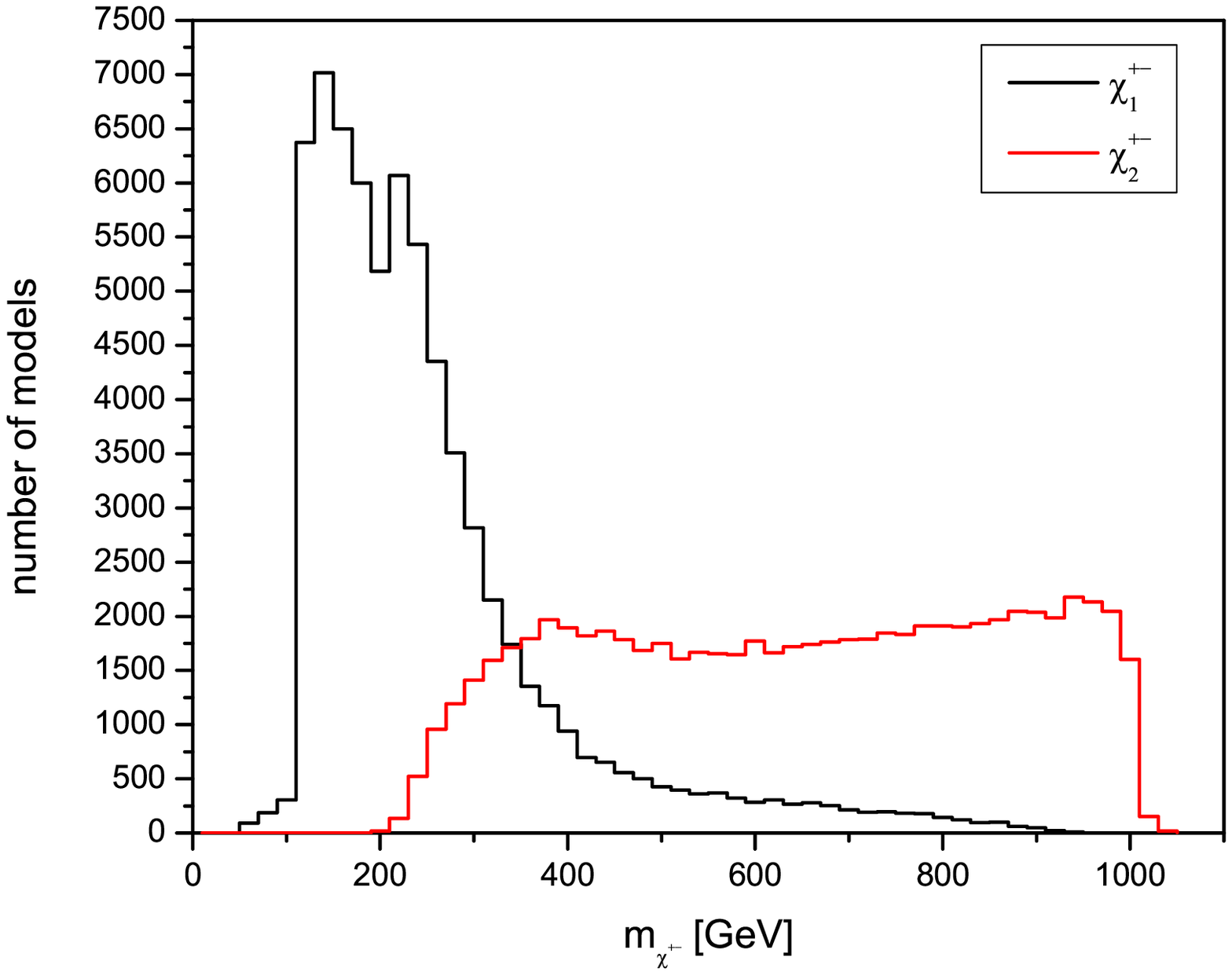}
\end{center}
\vspace{-1.5cm}
\caption{Distribution of neutralino masses (top panel) and
chargino masses (bottom panel) for our set of
  models.}
\label{fig1}
\end{figure}

Figure~\ref{fig1} presents a histogram of the masses of the four
neutralino and two chargino species in our models.  
The lightest neutralino is, of
course, the LSP.  The LSP mass lies between $100$ and $250$ GeV in over
$70\%$ of our models.  Generally models with a mostly Higgsino or Wino
LSP have a chargino with nearly the same mass as the LSP; as
sufficiently light
charginos would normally have been detected at LEP or the Tevatron, there
are fewer models with such LSPs with mass $\lsim 100$ GeV.

\begin{figure}[htbp]
\begin{center}
\includegraphics[width=8.5cm,angle=0]{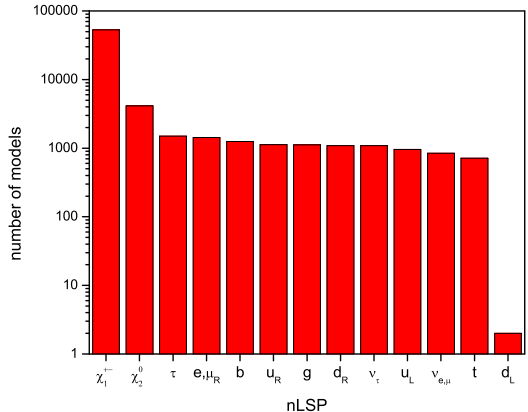}
\end{center}
\vspace{-1.5cm}
\caption{Number of models in which the nLSP is the given sparticle.}
\label{fig3}
\end{figure}

The identity of the nLSP is shown in Figure~\ref{fig3}.  The lightest chargino
is the nLSP in about $78\%$ of the models; this is due to many 
models having Wino or Higgsino LSPs, and the generally small mass
splitting between a mostly Wino or Higgsino neutralino and the
corresponding chargino.  The second lightest
neutralino is the nLSP $\sim 6\%$ of the time.  These will generally
be models with a dominantly Higgsino LSP.  Note also that while
neutralinos or charginos are the nLSP in the vast majority of cases,
there are 10 other sparticles each of which is the nLSP in $>1\%$ of
our models.  Scenarios in which
these sparticles are the nLSP may lead to interesting
signatures at the LHC\cite{Us ATLAS}.

\begin{figure}[htbp]
\begin{center}
\includegraphics[width=7.5cm,angle=0]{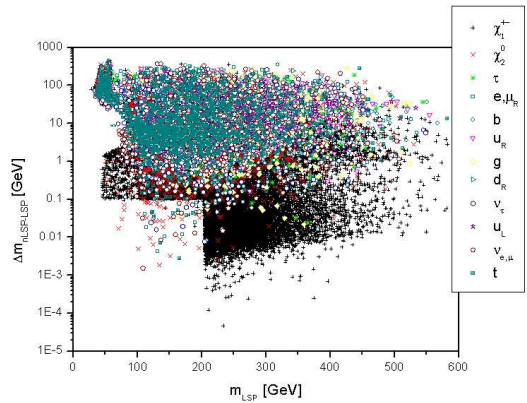}
\end{center}
\vspace{-1.5cm}
\caption{Mass splitting between nLSP and LSP versus LSP mass.  The identity
of the nLSP is shown as well.  (The LSP is always the lightest neutralino in
our set of models).}
\label{fig4}
\end{figure}

Figure~\ref{fig4} displays the LSP mass value as a function of the 
LSP-nLSP mass 
splitting, $\Delta m$,  our models for each identity of the LSP.  
It is interesting that these models have a smaller $\Delta m$ than
is often considered; $80\%$ of our models have $\Delta m < 10$ GeV,
$27\%$ have $\Delta m < 1$ GeV, and $3\%$ have $\Delta m < 10$ MeV.
As one can see from Figure~\ref{fig4}, this occurs largely, but not
exclusively, in models with a chargino nLSP. This is again due to
the many models where the LSP is nearly pure Wino or Higgsino.  

There are a number of interesting features in this figure.  The mostly
empty square region which appears on the lower left-hand side of
Figure~\ref{fig4} is due to the fact that models with chargino nLSPs
in this mass and $\Delta m$ range have been excluded by the Tevatron
stable chargino search.  Non-chargino nLSPs are not eliminated by this
search (\eg, the production cross section for sleptons in this range
is too small to be excluded by the Tevatron search).  It is perhaps
worth noting that a stable heavy charged particle search at the LHC,
corresponding to those done at the Tevatron, would be able to exclude
or discover the models with heavier chargino nLSPs and small values of
$\Delta m$ (corresponding to $\sim 12\%$ of our model set).

Another interesting feature in this figure is the bulge for $0.1$ GeV 
$\le \Delta m \lsim 2$ GeV and $m_{LSP}\lsim 100$ GeV.  
This region exists because
these values of $\Delta m$ are large enough that at LEP or the Tevatron,
the produced chargino would decay in the detector, but the resulting
charged tracks would be too soft to be observed.  The existence of such
a region shows the difficulty of making model independent statements
about sparticle masses or other SUSY observables.

We have seen that within our model set the nLSP can be almost any SUSY
particle and the corresponding $\Delta m$ can be small for these
cases. Thus 
specific models in our set describe qualitatively most of the 
conventional long-lived sparticle scenarios.
Long-lived stops or staus (as in GMSB), gluinos (as in
Split SUSY) as well as charginos (as in AMSB)
all occur in our sample.  
We also have long-lived neutralinos, as does GMSB, however 
these are the $\tilde \chi_2^0$ in our case.
In addition to models which, to some extent, correspond to these
well-studied scenarios, we also have models with long-lived selectrons,
sneutrinos and sbottoms.

\begin{figure}[htbp]
\begin{center}
\includegraphics[width=7.5cm,angle=0]{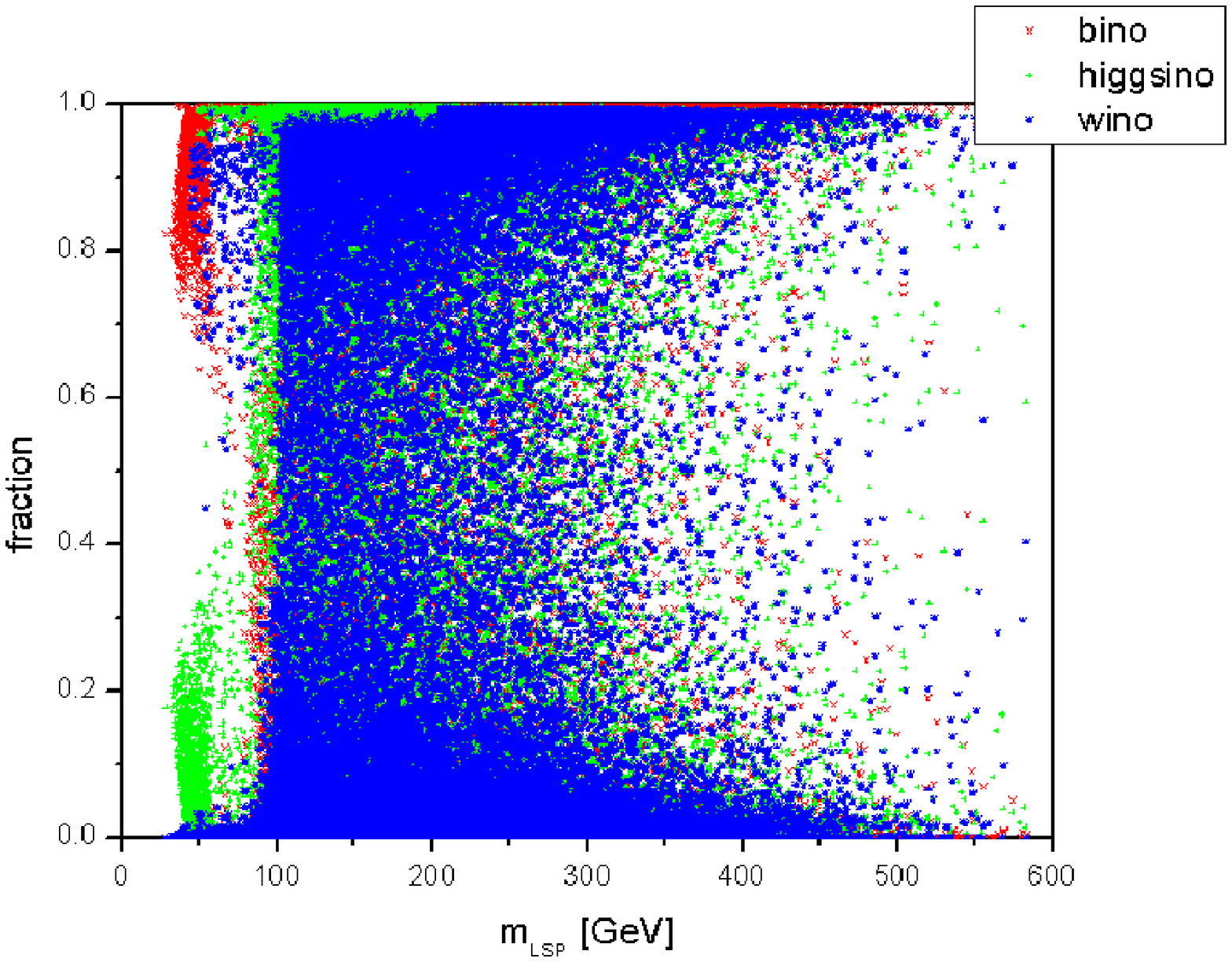}
\vspace*{-0.1cm}
\includegraphics[width=8.5cm,angle=0]{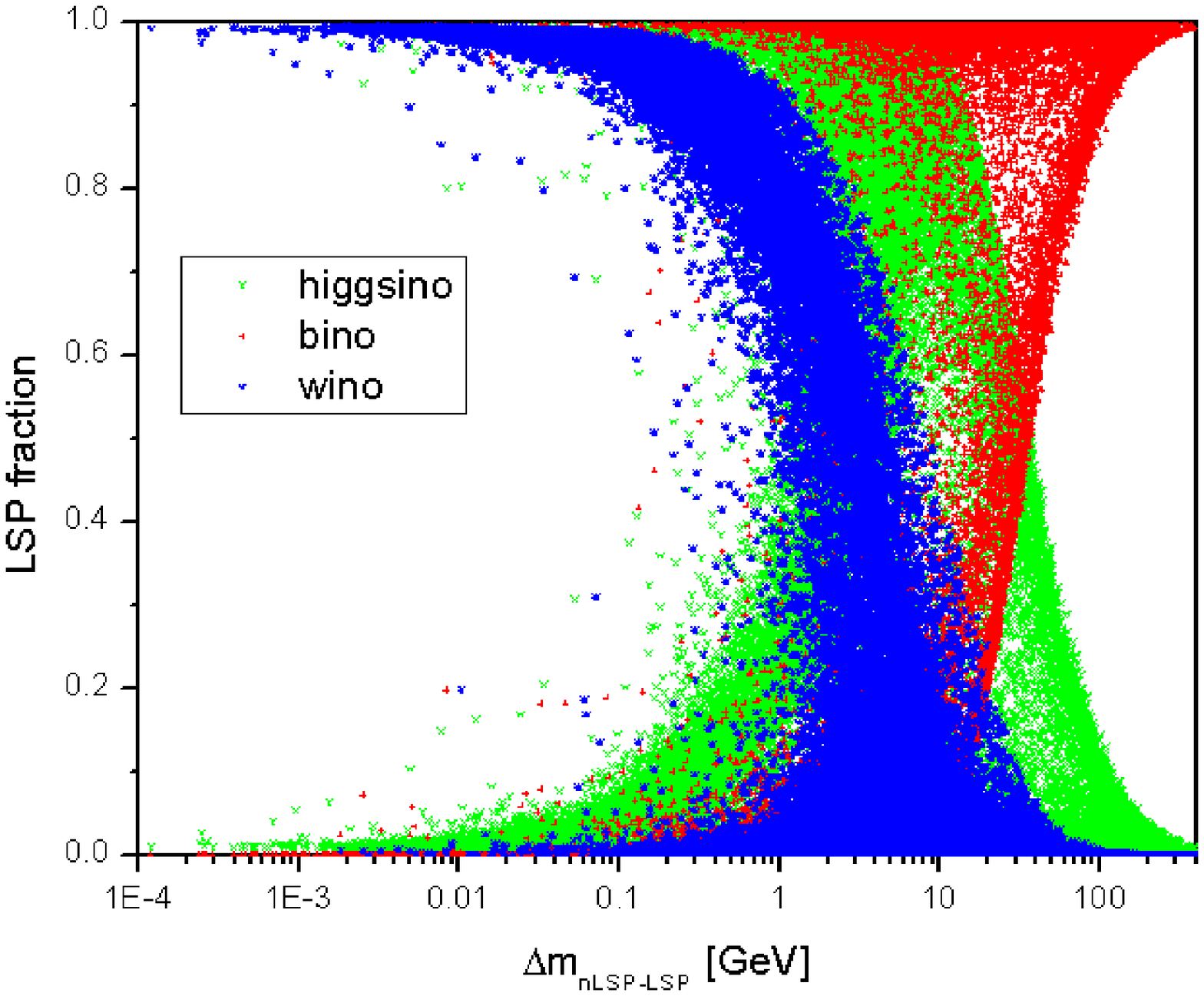}
\end{center}
\vspace{-1.5cm}
\caption{The distribution of LSP gaugino eigenstate types as a function
  of the LSP mass (top panel) or the LSP-nLSP mass difference (bottom panel).  
Note that each LSP corresponds to three points on
  this figure, one each for its Bino, Wino, and Higgsino fraction.}
\label{fig5}
\end{figure}

\begin{figure}[htbp]
\begin{center}
\includegraphics[width=8.5cm,angle=0]{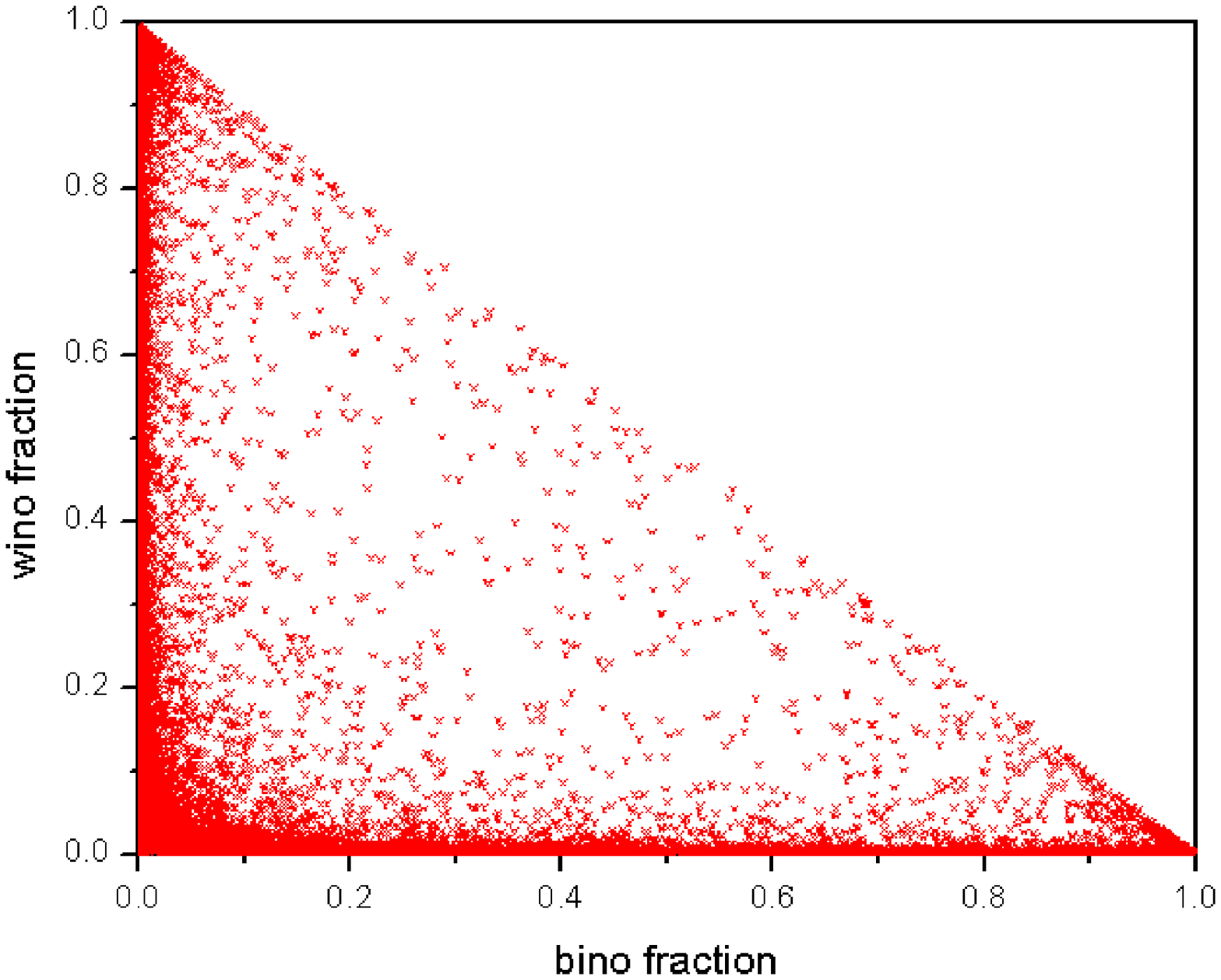}
\vspace*{-0.1cm}
\includegraphics[width=8.5cm,angle=0]{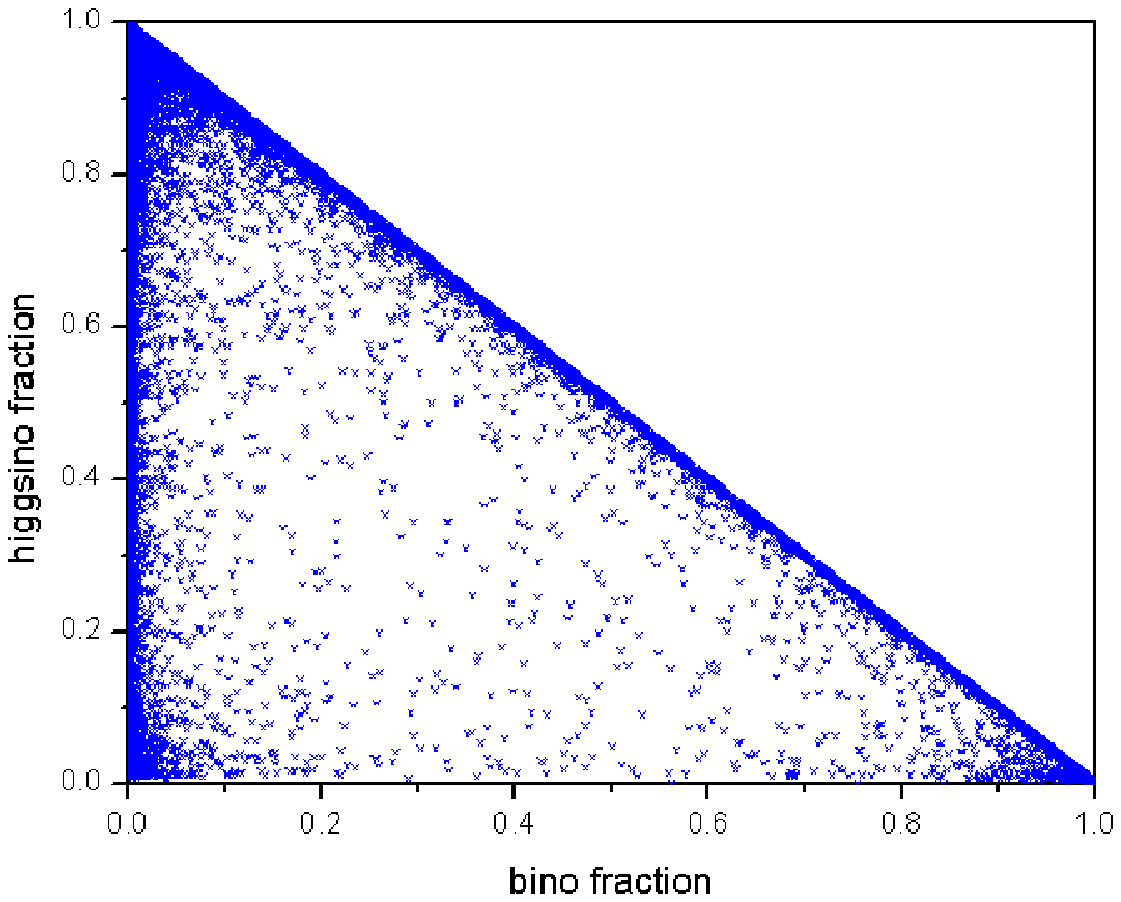}
\end{center}
\vspace{-1.5cm}
\caption{Wino/Higgsino/Bino content of the LSP in the case of flat
  priors.  Note that, as elsewhere in the paper, $|Z_{11}|^2$,
  $|Z_{12}|^2$, and $|Z_{13}|^2 + |Z_{14}|^2$, where $Z_{ij}$ is the
  neutralino mixing matrix, give
  the Bino, Wino, and Higgsino fractions respectively.
 }
\label{fig7}
\end{figure}

Figures~\ref{fig5},~and~\ref{fig7} display the gauge
eigenstate content of the LSPs in
our model set.  We note that most LSPs are relatively pure
eigenstates, with models where the LSP is Higgsino or mostly Higgsino
being by far the most common.  About one quarter of our models have
Wino or mostly Wino LSPs, while just over one-sixth have Bino or
mostly Bino LSPs. Within mSUGRA, the LSP is, in general, nearly purely
Bino; this suggests that most of our models are substantially
different from mSUGRA. 
We note that one would expect the LSP be a pure eigenstate
fairly often in a random scan of Lagrangian parameters, since if the
differences between $M_1,M_2,$ and $\mu$ are large compared to $M_Z$,
then the eigenstates of the mixing matrix will be essentially pure
gaugino and Higgsino states.  
 
\subsection{Relic Density}

\begin{figure}[htbp]
\begin{center}
\includegraphics[width=5.0cm,angle=-90]{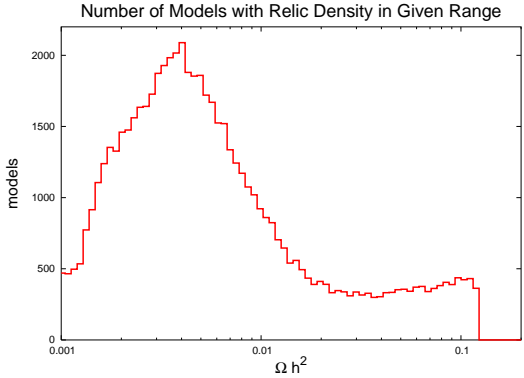}
\end{center}
\vspace{-0.7cm}
\caption{Distribution of $\Omega
h^2|_{\mathrm{LSP}}$ for our models.}
\label{fig8}
\end{figure}

\begin{figure}[t]
\begin{center}
\includegraphics[width=5.0cm,angle=-90]{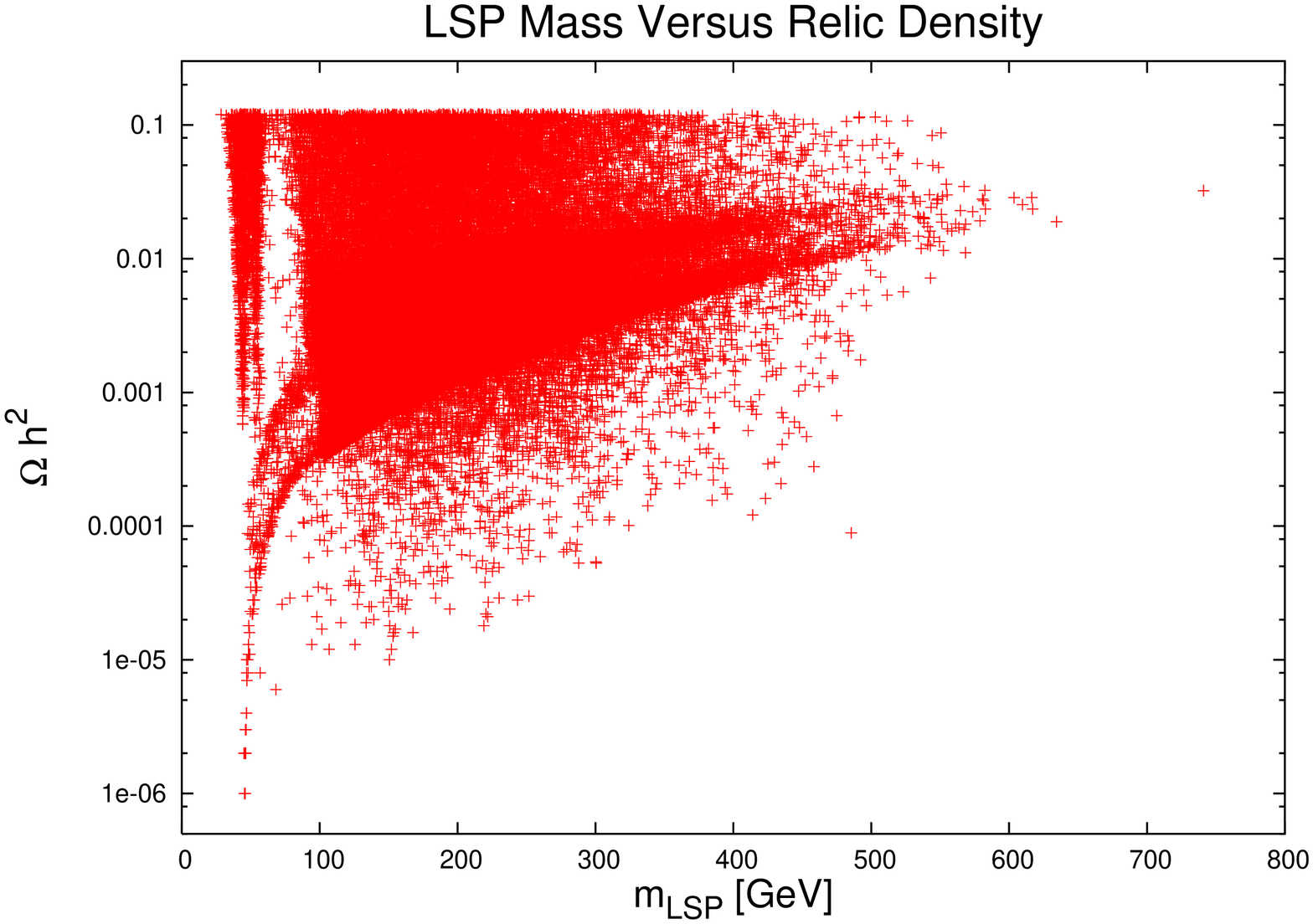}
\vspace*{-0.1cm}
\includegraphics[width=5.0cm,angle=-90]{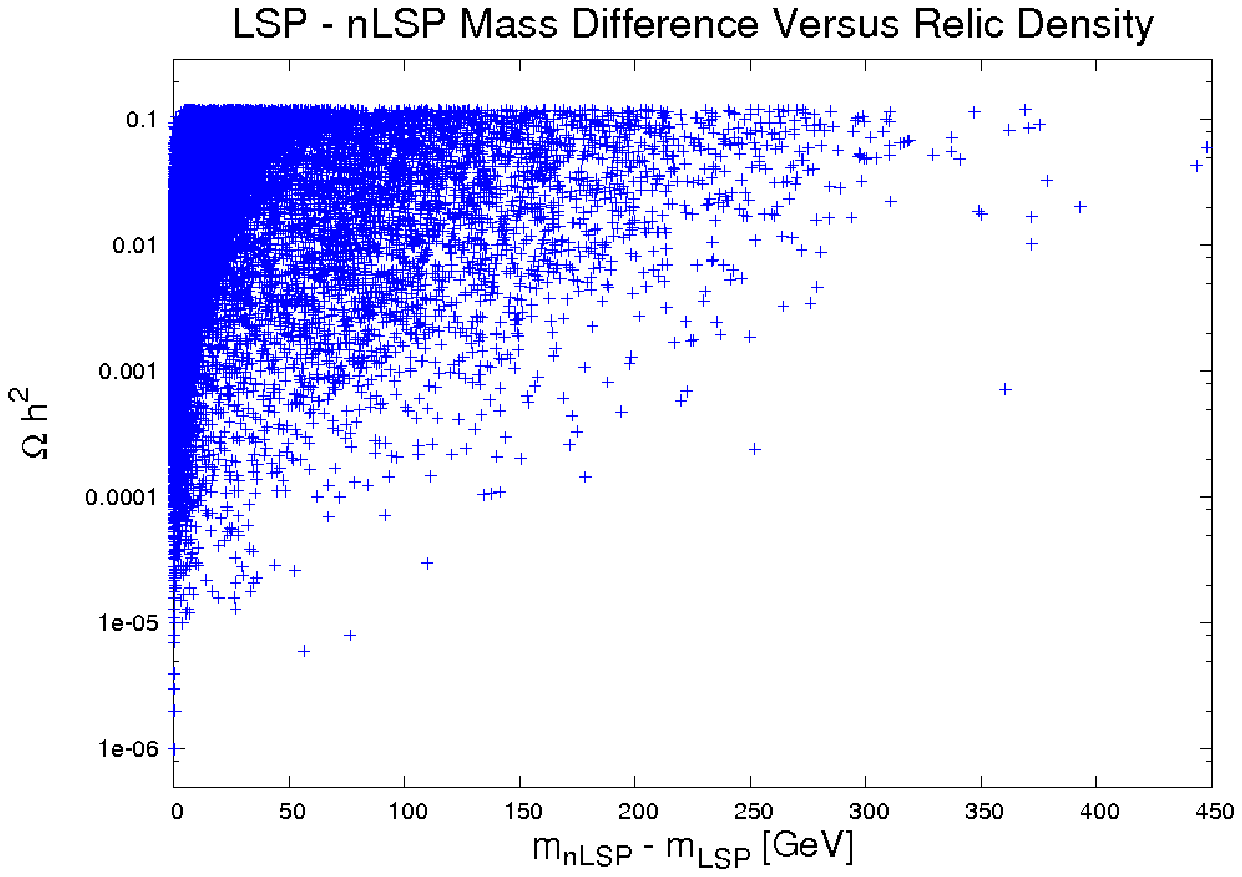}
\end{center}
\vspace{-0.7cm}
\caption{Distribution of $\Omega
h^2|_{\mathrm{LSP}}$ as a function of
  the LSP mass (top panel) or the LSP-nLSP mass splitting (bottom
panel.}
\label{fig9}
\end{figure}

We did not demand that the LSP, in any given model, account for
all of the dark matter, rather we required only that the LSP relic
density not be too large to be consistent with WMAP.  More
specifically, we employed $\Omega h^2|_{\mathrm{LSP}} < 0.121$.  
Figure~\ref{fig8} shows the
distribution of $\Omega h^2|_{\mathrm{LSP}}$ values predicted by our model
set.  Note that this distribution is peaked at small values of $\Omega
h^2|_{\mathrm{LSP}}$.  In particular, the mean value for this
quantity in our models is $\sim 0.012$. 
We note that the range of possible values of $\Omega
h^2|_{\mathrm{LSP}}$ is found to be much larger than those obtained by
analyses of specific SUSY breaking scenarios{\cite {big}}.
We display the predictions for $\Omega
h^2|_{\mathrm{LSP}}$ versus the LSP mass 
and versus the nLSP - LSP mass splitting in
Figure~\ref{fig9}.  Figure~\ref{fig9} makes it clear that
$\Omega h^2|_{\mathrm{LSP}}$ generally increases 
with the LSP mass, but a large
range of values for the relic density are possible at any given LSP mass.  
The empty region in 
Figure~\ref{fig9} where $\Omega h^2|_{\mathrm{LSP}} \approx 0.001 -
0.1$ and $m_{\mathrm{LSP}}\approx 50-100$ is due to the fact that, in
general, LSPs which are mostly Higgsino or Wino give lower values of
$\Omega h^2|_{\mathrm{LSP}}$, and, as noted above, there are fewer
Higgsino or Wino LSPs in this mass range.
This figure also shows that small mass differences can lead to large
dark matter annihilation rates.

\subsection{Direct Detection of Dark Matter}

As noted above, we calculate the spin-dependent and spin-independent
WIMP-nucleon cross sections using micrOMEGAs 2.21~\cite{MICROMEGAS}.
These data give the possible signatures in our
model set for experiments that search for WIMPs directly.  As these
experiments measure the product of WIMP-nucleon cross
sections with the local relic density, the cross section data presented in
the figures below are scaled by $\xi = \Omega
h^2|_{\mathrm{LSP}} / \Omega h^2|_{\mathrm{WMAP}}$.  To date,
  these experiments generally provide a more significant bound on the
  spin-independent cross section, and hence we will focus on those.

\begin{figure}[htbp]
\begin{center}
\includegraphics[width=5.5cm,angle=-90]{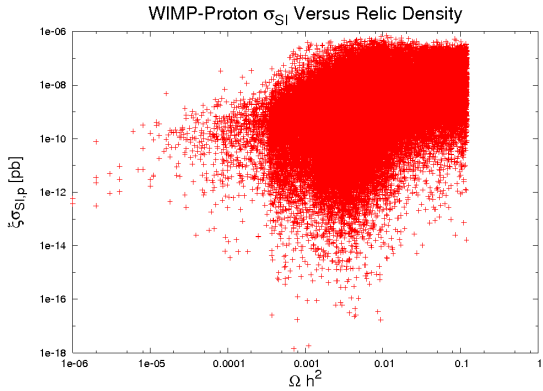}
\end{center}
\vspace{-0.7cm}
\caption{
Distribution of scaled WIMP-proton spin-independent cross section
versus the LSP contribution to relic density for our models.
}
\label{fig11}
\end{figure}

Figure~\ref{fig11} presents the distribution for the scaled WIMP-proton
spin-independent cross section versus relic density for our model
sample.  As one would expect, larger values of the cross section
 are generally found at larger values of $\Omega h^2|_{\mathrm{LSP}}$.  
However, even for relic densities close to the
  WMAP value, $\xi \sigma_{p,SI}$ is seen to vary by almost eight
  orders of magnitude.
These ranges for $\xi \sigma_{p,SI}$ are much larger than those from
mSUGRA as calculated, \eg, in~\cite{Barger:2008qd}.

\begin{figure}[htbp]
\begin{center}
\includegraphics[width=5.5cm,angle=-90]{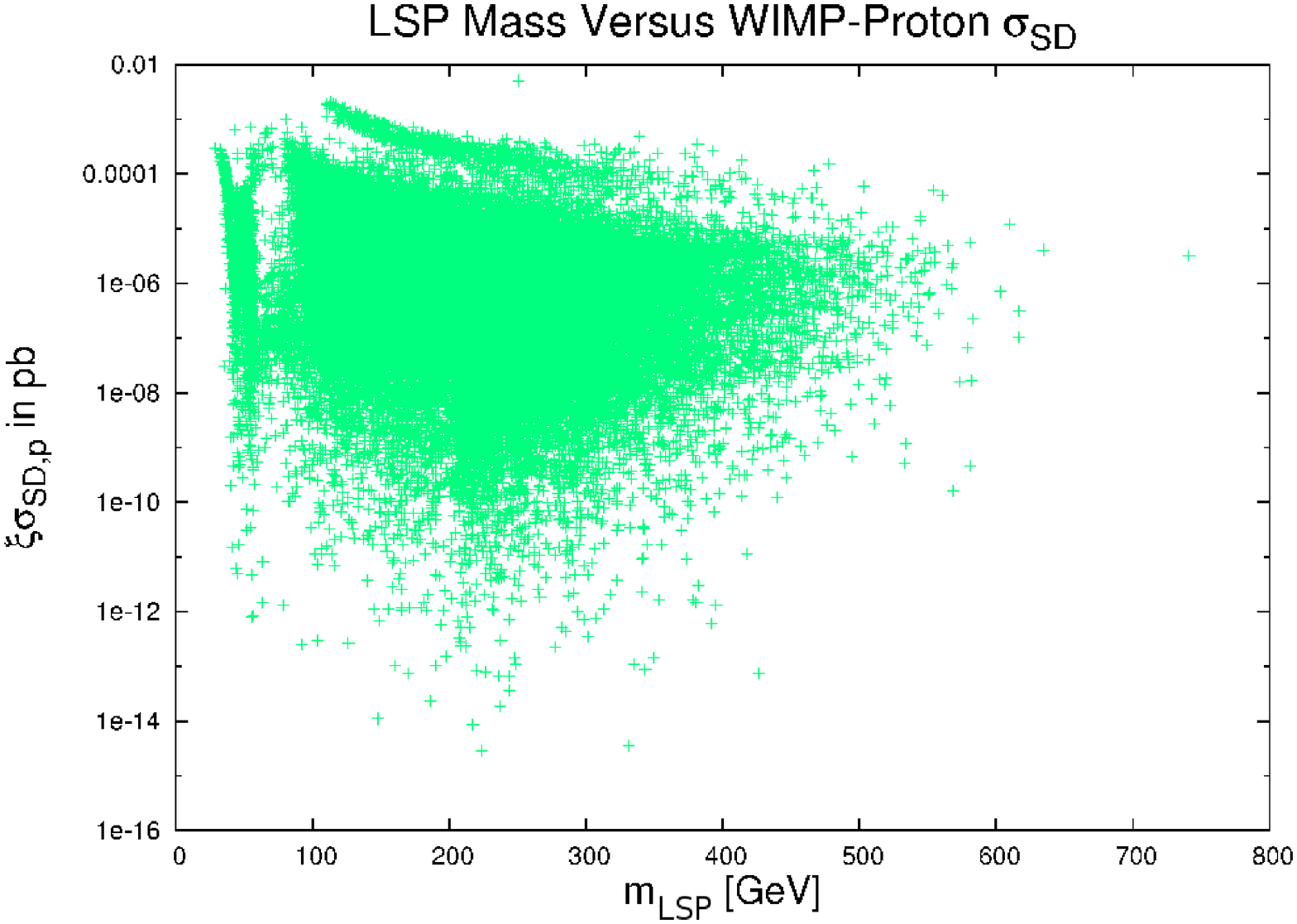}
\vspace*{1cm}
\includegraphics[width=5.0cm,angle=-90]{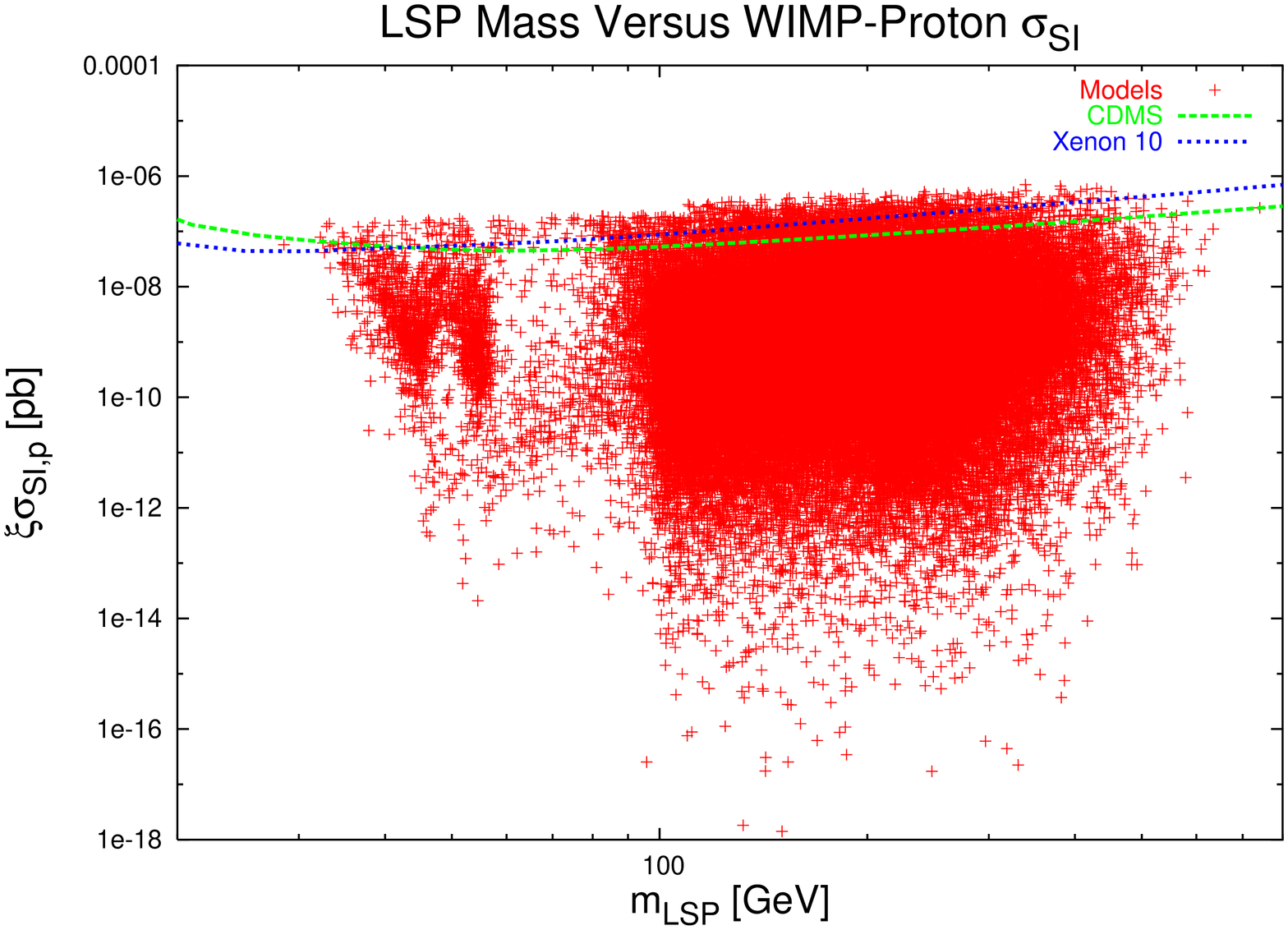}
\end{center}
\vspace{-1.5cm}
\caption{
Distributions of scaled WIMP-proton spin-dependent cross section and
spin-independent cross sections versus LSP mass in our models.  In the
spin-independent panel, the CDMS and Xenon10 bounds are shown.
 }
\label{fig12}
\end{figure}

Figure~\ref{fig12} shows the scaled WIMP-proton spin-dependent and
spin-independent cross sections as a function of the LSP mass.  The
constraints from XENON10\cite{XENON10} and CDMS\cite{CDMS} are also
displayed.  
As noted above, to take the uncertainties in the theoretical
calculations of the WIMP-nucleon cross section into account, we
allowed for a factor of 4 uncertainty in the calculation of the WIMP-nucleon
cross section.
Table 3 in Ref.{\cite {us2}} gives the
fraction of models that would be excluded if the combined
CDMS/XENON10 cross section limit
were improved by an overall scaling factor. 
Note that our inclusion of the theoretical uncertainties does not
significantly modify the size of our model sample.

We find that the range of values obtained for these cross sections
covers the entire
region in cross section/ LSP space that is anticipated from different
types of Beyond the Standard Model theories in the above reference.
This suggests that we cannot use direct detection experiments to
distinguish between \eg~SUSY versus Little Higgs versus Universal
Extra Dimensions dark matter candidates in the absence of other data.

\begin{figure}[htbp]
\begin{center}
\includegraphics[width=5.5cm,angle=-90]{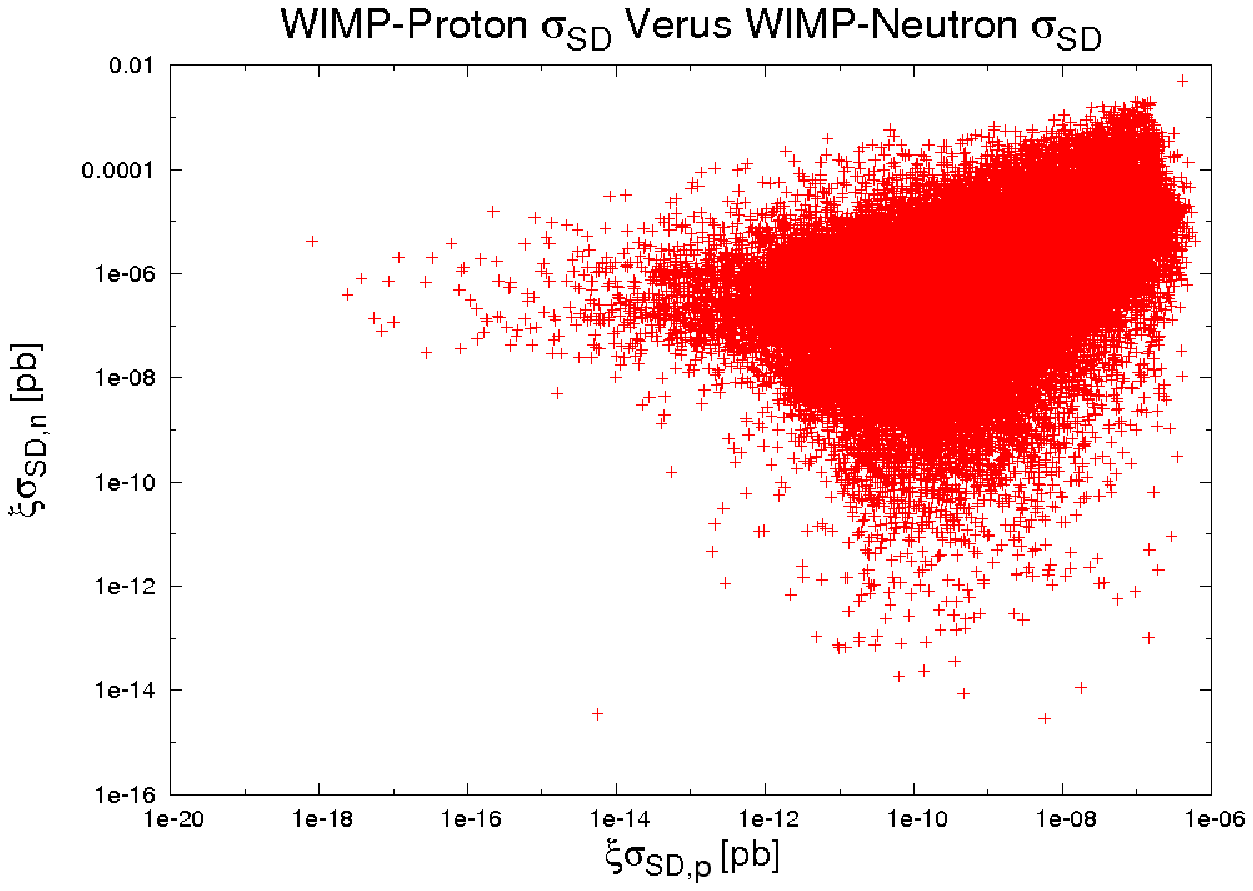}
\vspace*{0.1cm}
\includegraphics[width=5.5cm,angle=-90]{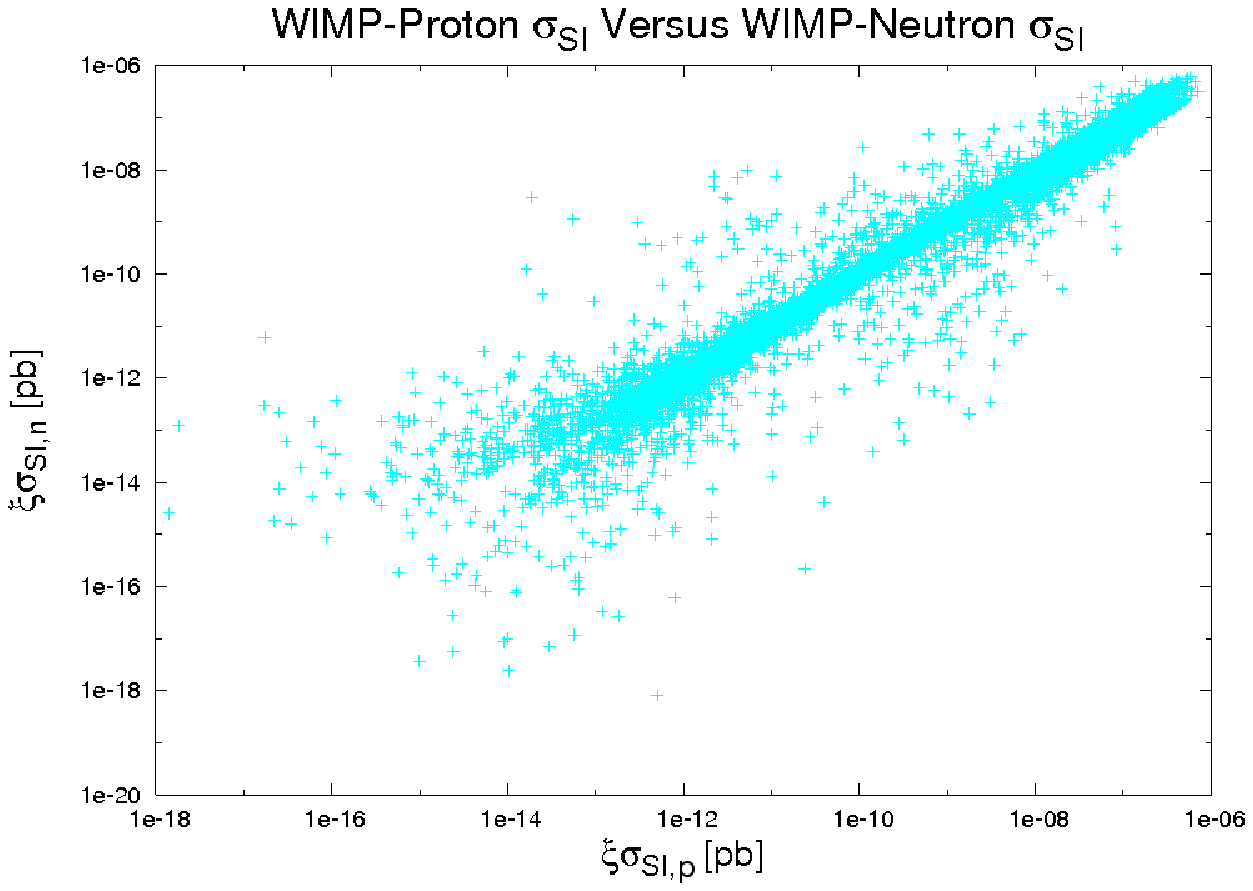}
\end{center}
\vspace{-1.3cm}
\caption{
Comparison of the WIMP-neutron and WIMP-proton cross sections.  The
spin-dependent(independent) cross sections are shown in the 
top(bottom) panel.
 }
\label{fig13}
\end{figure}

In Figure~\ref{fig13}, we compare the WIMP-proton and WIMP-neutron cross
sections in the spin-dependent and spin-independent cases.  The
spin-independent cross sections are seen to be fairly isospin independent;
this is not the case, however, for the spin-dependent cross sections.


\subsection{Indirect Detection of Dark Matter}

The PAMELA collaboration has recently claimed an excess in the ratio
of cosmic ray positrons to electrons observed at energies
$\gsim 10$ GeV\cite{Adriani:2008zr}.
Here we employ DarkSUSY 5.0.4\cite{DarkSUSY} to calculate this ratio for
our model sample and compare these results with the PAMELA data.

In general, for a thermal relic dark matter candidate to reproduce
the PAMELA data, its signal rate must be multiplied by a boost 
factor\cite{boost}.  
In nature, such a boost factor could
result from, \eg, a local overdensity.  The boost factor in that case
would be the square of the ratio between the density of dark matter in
the region from which one is sensitive to cosmic ray positrons and
electrons to the universe as a whole.

In our analysis, we use four propagation models which are present 
in darkSUSY: the
model of Baltz and Edsj\"{o}\cite{Baltz:1998xv}, that of Kamionkowski
and Turner\cite{Kamionkowski:1990ty}, that of Moskalenko and
Strong\cite{Moskalenko:1999sb}, as well as
GALPROP\cite{GALPROP}.  These are referred to in
the following figures as ``BE'',``KT'',``MS'', and ``GAL'',
respectively.
Interestingly, we find that the extent to which the
positron/electron flux ratio predicted by our models matches the PAMELA
data can be highly sensitive to the choice of propagation model
parameters.  We will explore this further in future work\cite{Us
  DM}. 

\begin{figure}[htbp]
\begin{center}
\includegraphics[width=5.0cm,angle=-90]{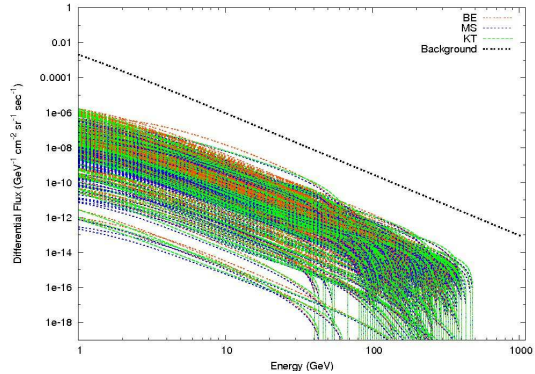}
\end{center}
\vspace{-1.0cm}
\caption{
Expected flux spectrum of positrons from neutralino annihilation in
the halo for 500 randomly selected models.  For each model, there are
three curves, one for each of three propagation models (as shown in the legend 
and defined in the text). 
The dotted black line is the expected background of positrons from
non-SUSY processes.
}
\label{fig14}
\end{figure}

\begin{figure}[htbp]
\begin{center}
\includegraphics[width=5.0cm,angle=-90]{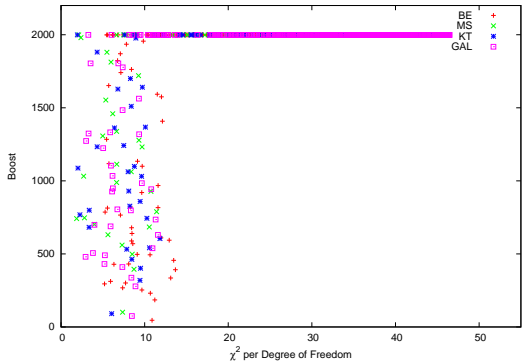}
\end{center}
\vspace{-1.0cm}
\caption{
The distribution of $\chi^2$ per degree of freedom versus the choice
of boost factor that minimized this quantity for 500 randomly selected
pMSSM models in our model set.  These quantities have
been determined for each of four propagation models.  Only
boost factors less than $2000$ were considered; this explains the
large number of models for which the $\chi^2$-minimizing boost was $2000$. 
}
\label{fig15}
\end{figure}

\begin{figure}[htbp]
\begin{center}
\includegraphics[width=5.0cm,angle=-90]{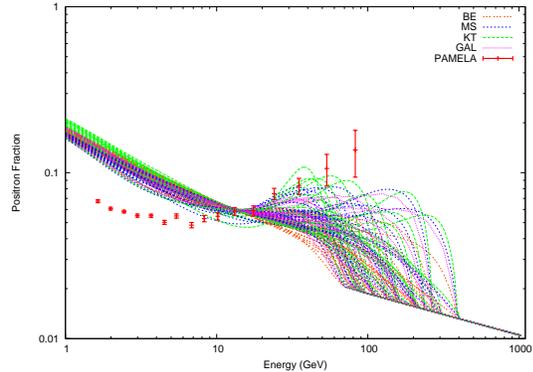}
\end{center}
\vspace{-1.0cm}
\caption{
Positron/ electron flux ratio versus energy for the pMSSM models
for which the $\chi^2$ per degree of freedom with the
$\chi^2$-maximizing boost was less than $10.0$ for three of the four
propagation models.  Curves are shown for all four propagation models.
}
\label{fig16}
\end{figure}

The differential positron flux as a function of energy for 
a random sample of 500 models from our set are shown  in
Figure~\ref{fig14}.  Here we assume a boost factor of $1$; the
normalization of the curves takes into account the fact that for many
of these models  $\Omega h^2|_{\mathrm{LSP}} < \Omega_{\mathrm{WMAP}}$.
  
We next determine how well the predicted positron fluxes for these
models agree with the PAMELA data, allowing for the possibility of a
boost factor.  To do this, we find the value for the boost factor
(with the restriction that it be $< 2000$)
which minimizes the $\chi^2$
for the fit of each model's prediction to the PAMELA data.  In
calculating the $\chi^2$, we consider only the seven highest energy
bins, as at lower energies solar modulation is expected to play a major
role\cite{Adriani:2008zr}.  Figure~\ref{fig15} shows the $\chi^2$ and
corresponding boost factor for these 500 random models.
Note that there are four data points for each model, so there
are actually 2000 data points in this figure.  
We then display the positron to electron flux ratio, for the models
with a low value of $\chi^2$, as a function of energy in
Figure~\ref{fig16}, and note the reasonable agreement with the data
for some models.

\begin{figure}[htbp]
\begin{center}
\includegraphics[width=5.0cm,angle=-90]{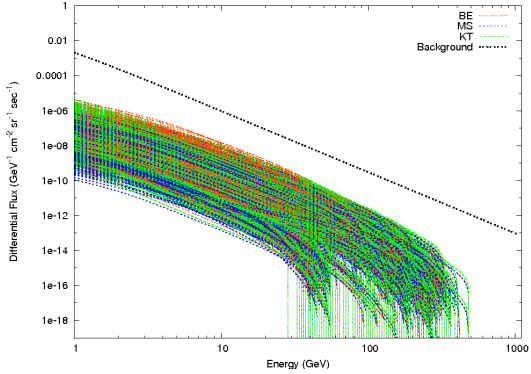}
\end{center}
\vspace{-1.0cm}
\caption{
Expected flux spectrum of positrons from neutralino annihilation in
the halo for 500 randomly selected models for which
$\Omega h^2|_\mathrm{WMAP} \ge \Omega h^2|_\mathrm{LSP}>0.10$.  For each model, there are
three curves, one for each of three propagation models (as shown in the legend). 
The dotted black line is the expected background of positrons from
non-SUSY processes.
}
\label{fig17}
\end{figure}

\begin{figure}[htbp]
\begin{center}
\includegraphics[width=5.0cm,angle=-90]{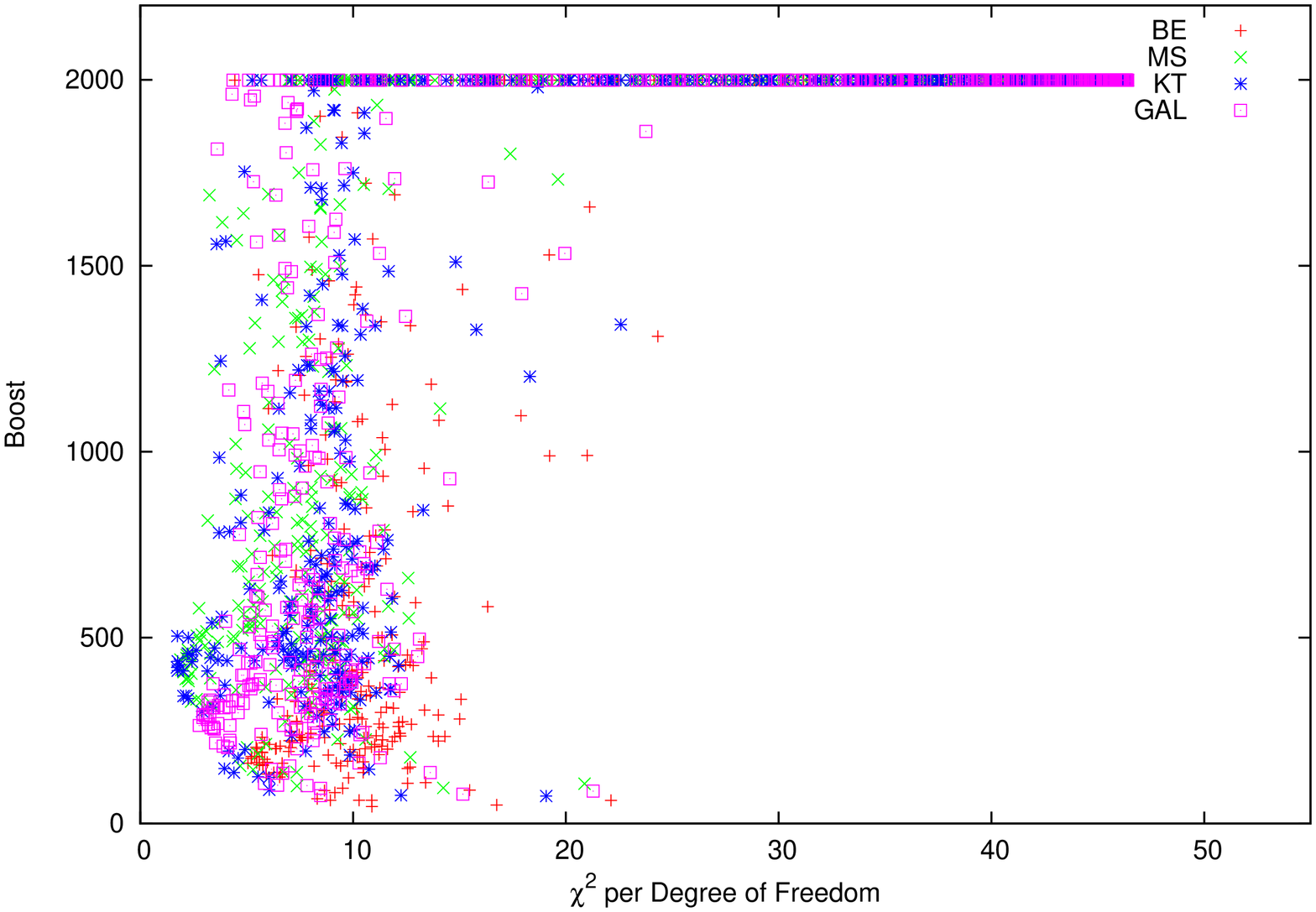}
\end{center}
\vspace{-1.0cm}
\caption{
The distribution of $\chi^2$ per degree of freedom versus the choice
of boost factor that minimized this quantity for 500 randomly selected
pMSSM models with $\Omega h^2|_\mathrm{WMAP} \ge \Omega h^2|_{\mathrm{LSP}}>0.10$.  
These quantities have
been determined for each of four propagation models.  Only
boost factors less than $2000$ were considered; this explains the
large number of models for which the $\chi^2$-minimizing boost was $2000$. 
}
\label{fig18}
\end{figure}

\begin{figure}[htbp]
\begin{center}
\includegraphics[width=5.0cm,angle=-90]{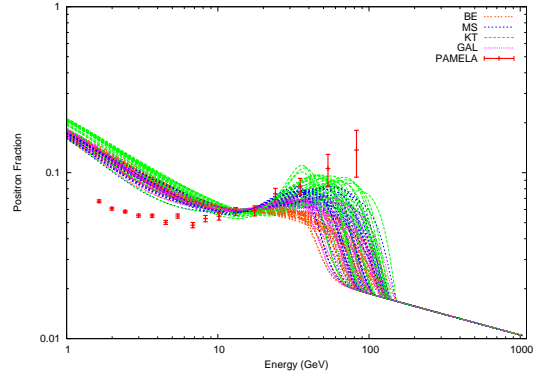}
\end{center}
\vspace{-1.0cm}
\caption{
Positron/ electron flux ratio versus energy curves for pMSSM models
for which the $\chi^2$ per degree of freedom with the
$\chi^2$-maximizing boost was less than $5.0$ for three of the four
propagation models.  Curves are shown for all four propagation models.
}
\label{fig19}
\end{figure}

Since the flux from WIMP annihilation scales as $(\Omega
h^2|_{\mathrm{LSP}}/\Omega h^2|_{\mathrm{WMAP}})^2$, we might expect to
improve the match to the PAMELA data using models from our sample
for which
$\Omega h^2|_{\mathrm{LSP}}\approx \Omega h^2|_{\mathrm{WMAP}}$.  To
test this, we examine the predicted positron flux for 500 random models
with $\Omega h^2|_{\mathrm{LSP}} > 0.100$;
these fluxes are shown in Figure~\ref{fig17} with no boost factor.  
We then again find the boost
factor that minimizes the $\chi^2$ of the positron to electron flux
ratios with respect to the seven highest energy PAMELA bins;
these are shown in Figure~\ref{fig18}.  Here, we note that there are many more
models for which the $\chi^2$-minimizing value for the boost factor is $<
2000$ and there are many more points for which the $\chi^2$ value is low.  
The positron to electron flux ratios for these models, including the
boost factor, are shown in Figure~\ref{fig19}.

It appears that some of our models do a reasonably good job
of fitting the PAMELA positron data, especially in the case where $\Omega
h^2|_{\mathrm{LSP}}$ lies fairly close to the WMAP value.  
For most models, describing the PAMELA data requires large boost
factors, however this is
also a fairly generic feature of attempts to explain PAMELA and ATIC
data in terms of WIMP annihilation\cite{boost}.  There are however,
many models which give relatively low $\chi^2$ per degree of freedom
in the fit to the data with relatively small boost factors.  We will
study this further in future work \cite{Us DM}.  A study of the
corresponding predictions for the the cosmic ray anti-proton flux is also
underway.

\section{Conclusions}

We have generated a large set of points in parameter space (which we call
``models'') for the 19-parameter CP-conserving pMSSM, where MFV has been
assumed.  We subjected these models to numerous experimental and
theoretical constraints to obtain a set of $\sim 68$~K models which
are consistent with existing data.  We attempted to be
somewhat conservative in our implementation of these constraints; in
particular we only demanded that the relic density of the LSP not
be greater than the measured value of $\Omega h^2$ for non-baryonic
dark matter, rather than assuming that the LSP must account for the
\textit{entire} observed relic density.
 
Examining the properties of the neutralinos in these models, we
find
that many are relatively pure gauge eigenstates with Higgsinos being
the most common, followed by Winos.  
The relative prevalence of Higgsino and Wino LSPs leads many
of our models to have a chargino as nLSP, often with a relatively
small mass splitting between this nLSP and the LSP; this has important
consequences in both collider and astroparticle phenomenology.

We find that, in general, the LSP in our models provides a relatively
small ($\sim 4\%$) contribution to the dark matter, however there is a
long tail to this distribution and a substantial number of models for
which the LSP makes up all or most of the dark matter.  Typically
these neutralinos are mostly Binos.  

Examining the signatures of our models in direct and indirect dark
matter detection experiments, we find a wide range of signatures for
both cases.  In particular, we find a much larger range of WIMP-nucleon
cross sections than is found in any particular model of SUSY-breaking.  As
these cross sections also enter the regions of parameter space
suggested by non-SUSY models, it appears that the discovery of WIMPs
in direct detection experiments might not be sufficient to determine
the correct model of the underlying physics.  As a first look at the
signatures of these models in indirect detection experiments, we
examined whether our models could explain the PAMELA excess in
the positron to electron ratio at high energies.  We find that there
are models which fit the PAMELA data rather well, and some of these
have significantly smaller boost factors than generally assumed for a
thermal relic.

The study of the pMSSM presents exciting new possibilities for SUSY
phenomenology.  The next few
years will hopefully see important discoveries both in colliders and
in satellite or ground-based astrophysical experiments.  It is
important that we follow the data and not our existing prejudices;
hopefully this sort of relatively model-independent approach to
collider and astrophysical phenomenology can be useful in this regard.

%
\def\MPL #1 #2 #3 {Mod. Phys. Lett. {\bf#1},\ #2 (#3)}
\def\NPB #1 #2 #3 {Nucl. Phys. {\bf#1},\ #2 (#3)}
\def\PLB #1 #2 #3 {Phys. Lett. {\bf#1},\ #2 (#3)}
\def\PR #1 #2 #3 {Phys. Rep. {\bf#1},\ #2 (#3)}
\def\PRD #1 #2 #3 {Phys. Rev. {\bf#1},\ #2 (#3)}
\def\PRL #1 #2 #3 {Phys. Rev. Lett. {\bf#1},\ #2 (#3)}
\def\RMP #1 #2 #3 {Rev. Mod. Phys. {\bf#1},\ #2 (#3)}
\def\NIM #1 #2 #3 {Nuc. Inst. Meth. {\bf#1},\ #2 (#3)}
\def\ZPC #1 #2 #3 {Z. Phys. {\bf#1},\ #2 (#3)}
\def\EJPC #1 #2 #3 {E. Phys. J. {\bf#1},\ #2 (#3)}
\def\IJMP #1 #2 #3 {Int. J. Mod. Phys. {\bf#1},\ #2 (#3)}
\def\JHEP #1 #2 #3 {J. High En. Phys. {\bf#1},\ #2 (#3)}


\begin{thebibliography}{99}


\bibitem{mfv}
For a review, see, 
G.~D'Ambrosio, G.~F.~Giudice, G.~Isidori and A.~Strumia,
  Nucl.\ Phys.\  B {\bf 645}, 155 (2002)
  [arXiv:hep-ph/0207036].

\bibitem{Djouadi:2002ze} 
  A.~Djouadi, J.~L.~Kneur and G.~Moultaka,
  Comput.\ Phys.\ Commun.\  {\bf 176}, 426 (2007)
  [arXiv:hep-ph/0211331].

\bibitem{Berger:2008cq}
  C.~F.~Berger, J.~S.~Gainer, J.~L.~Hewett and T.~G.~Rizzo,
  JHEP {\bf 0902}, 023 (2009)
  [arXiv:0812.0980 [hep-ph]].

\bibitem{us2}
  R.~C.~Cotta, J.~S.~Gainer, J.~L.~Hewett and T.~G.~Rizzo,
  arXiv:0903.4409 [hep-ph].


\bibitem{MICROMEGAS}
  G.~Belanger, F.~Boudjema, A.~Pukhov and A.~Semenov,
  Comput.\ Phys.\ Commun.\  {\bf 177} (2007) 894; 
  arXiv:0803.2360 [hep-ph];
 Comput.\ Phys.\ Commun.\  {\bf 149}, 103 (2002)
 [arXiv:hep-ph/0112278];
 Comput.\ Phys.\ Commun.\  {\bf 174}, 577 (2006)
 [arXiv:hep-ph/0405253];
 Comput.\ Phys.\ Commun.\  {\bf 176}, 367 (2007)
 [arXiv:hep-ph/0607059] and 
 arXiv:0803.2360 [hep-ph].

\bibitem{gino}
  G.~Isidori and P.~Paradisi,
  Phys.\ Lett.\  B {\bf 639}, 499 (2006)
  [arXiv:hep-ph/0605012].

\bibitem{ems}
  D.~Eriksson, F.~Mahmoudi and O.~Stal,
  arXiv:0808.3551 [hep-ph].

\bibitem{Bennett:2006fi}
  G.~W.~Bennett {\it et al.}  [Muon G-2 Collaboration],
  Phys.\ Rev.\  D {\bf 73}, 072003 (2006)
  [arXiv:hep-ex/0602035].

\bibitem{mesonmix}
For an introduction, see, 
  J.~S.~Hagelin, S.~Kelley and T.~Tanaka,
  Nucl.\ Phys.\  B {\bf 415}, 293 (1994);
  F.~Gabbiani {\it et al.},
  Nucl.\ Phys.\  B {\bf 477}, 321 (1996)
  [arXiv:hep-ph/9604387].



\bibitem{lepstable}
  G.~Benelli,
  ``Search for stable and long lived heavy charged particles in electron
  positron collisions at center of mass energies from 130-GeV to 209-GeV with
  the OPAL detector at LEP,'' UMI-31-09638, 2003. 126pp.


\bibitem{LEPEWWG}
LEP Electroweak Working Group, http://www.cern.ch/LEPEWWG.


\bibitem{ALEPH}
  R.~Barate {\it et al.}  [ALEPH Collaboration],
  Phys.\ Lett.\  B {\bf 469}, 303 (1999).

\bibitem{bbb}
  A.~C.~Kraan,
  arXiv:hep-ex/0305051.

\bibitem{LEPSUSY}
LEP SUSY Working Group, http://lepsusy.web.cern.ch/lepsusy/.

\bibitem{LEPHIGGS}
LEP Higgs Working Group, http://lephiggs.web.cern.ch/LEPHIGGS/
www/Welcome.html.

\bibitem{SUSYHIT}
  A.~Djouadi, M.~M.~Muhlleitner and M.~Spira,
  Acta Phys.\ Polon.\  B {\bf 38}, 635 (2007)
  [arXiv:hep-ph/0609292];
http://lappweb.in2p3.fr/\~{}tmuehlleitner/SUSY-HIT/.

\bibitem{uncertain}
For a review, see, for example, 
  S.~Heinemeyer, W.~Hollik and G.~Weiglein,
  Phys.\ Rept.\  {\bf 425}, 265 (2006)
  [arXiv:hep-ph/0412214].

\bibitem{domet}
  V.~M.~Abazov {\it et al.}  [D0 Collaboration],
  Phys.\ Lett.\  B {\bf 660}, 449 (2008)
  [arXiv:0712.3805 [hep-ex]].

\bibitem{PYTHIA}
  T.~Sjostrand, S.~Mrenna and P.~Skands,
  JHEP {\bf 0605}, 026 (2006)
  [arXiv:hep-ph/0603175].

\bibitem{PGS}
J. Conway, PGS4, Pretty Good detector Simulation, \\
http://www.physics.ucdavis.edu/\~{}conway/
research/software/pgs/pgs.html.

\bibitem{PROSPINO}
  W.~Beenakker, R.~Hopker and M.~Spira,
  arXiv:hep-ph/9611232;
 W.~Beenakker, R.~Hopker, M.~Spira and P.~M.~Zerwas,
 Nucl.\ Phys.\  B {\bf 492}, 51 (1997)
 [arXiv:hep-ph/9610490].
See also, 
 http://www.ph.ed.ac.uk/\~{}tplehn/prospino/. 

\bibitem{cdftrilepton}
  T.~Aaltonen {\it et al.}  [CDF Collaboration],
  arXiv:0808.2446 [hep-ex].


\bibitem{dostable}
  V.~M.~Abazov {\it et al.}  [D0 Collaboration],
  arXiv:0809.4472 [hep-ex].

\bibitem{stops}
  T.~Aaltonen {\it et al.}  [CDF Collaboration],
  Phys.\ Rev.\ Lett.\  {\bf 101}, 071802 (2008)
  [arXiv:0802.3887 [hep-ex]] and  
  T.~Aaltonen {\it et al.}  [CDF Collaboration],
  Phys.\ Rev.\  D {\bf 76}, 072010 (2007)
  [arXiv:0707.2567 [hep-ex]];
  V.~M.~Abazov {\it et al.}  [D0 Collaboration],
  Phys.\ Rev.\ Lett.\  {\bf 97}, 171806 (2006)
  [arXiv:hep-ex/0608013] and
  Phys.\ Lett.\  B {\bf 665}, 1 (2008)
  [arXiv:0803.2263 [hep-ex]]; 


\bibitem{Komatsu:2008hk}
  E.~Komatsu {\it et al.}  [WMAP Collaboration],
  arXiv:0803.0547 [astro-ph].

\bibitem{dmsearch}
For a recent overview see 
  J.~D.~Vergados,
  Lect.\ Notes Phys.\  {\bf 720}, 69 (2007)
  [arXiv:hep-ph/0601064].

 \bibitem{XENON10}
  U.~Oberlack  [XENON Collaboration],
  J.\ Phys.\ Conf.\ Ser.\  {\bf 110}, 062020 (2008).

\bibitem{CDMS}
  Z.~Ahmed {\it et al.}  [CDMS Collaboration],
  arXiv:0802.3530 [astro-ph].

\bibitem{CRESST}
  M.~Bravin {\it et al.}  [CRESST-Collaboration],
  Astropart.\ Phys.\  {\bf 12}, 107 (1999)
  [arXiv:hep-ex/9904005].

\bibitem{DAMA}
  R.~Bernabei {\it et al.}  [DAMA Collaboration],
  Phys.\ Lett.\  B {\bf 480}, 23 (2000).

\bibitem{Us ATLAS}
J.~A.~Conley, J.~S.~Gainer, J.~L.~Hewett, and T.~G.~Rizzo.  
In preparation.

    \bibitem{big}
There have been many such a analyses; a few recent representative examples are: 
  B.~C.~Allanach, M.~J.~Dolan and A.~M.~Weber,
  JHEP {\bf 0808}, 105 (2008)
  [arXiv:0806.1184 [hep-ph]]; 
  O.~Buchmueller {\it et al.},
  arXiv:0808.4128 [hep-ph];
  R.~Trotta, F.~Feroz, M.~P.~Hobson, L.~Roszkowski and R.~R.~de Austri,
  arXiv:0809.3792 [hep-ph];
  U.~Chattopadhyay and D.~Das,
  arXiv:0809.4065 [hep-ph];
  S.~Heinemeyer, X.~Miao, S.~Su and G.~Weiglein,
  arXiv:0805.2359 [hep-ph];
  J.~R.~Ellis, S.~Heinemeyer, K.~A.~Olive, A.~M.~Weber and G.~Weiglein,
  JHEP {\bf 0708}, 083 (2007)
  [arXiv:0706.0652 [hep-ph]];

\bibitem{Barger:2008qd}
  V.~Barger, W.~Y.~Keung and G.~Shaughnessy,
  arXiv:0806.1962 [hep-ph].

\bibitem{Adriani:2008zr}
  O.~Adriani {\it et al.},
  arXiv:0810.4995 [astro-ph].

\bibitem{DarkSUSY}
  P.~Gondolo, J.~Edsj\"{o}, P.~Ullio, L.~Bergstrom, M.~Schelke and E.~A.~Baltz,
  JCAP {\bf 0407}, 008 (2004)
  [arXiv:astro-ph/0406204].
  P. Gondolo, J. Edsjö, P. Ullio, L. Bergström, M. Schelke,
  E.A. Baltz, T. Bringmann and G. Duda,
  http://www.physto.se/\~{}edsjo/darksusy.


  \bibitem{boost}
    This is a fairly generic result, see for example:
  M.~Cirelli, M.~Kadastik, M.~Raidal and A.~Strumia,
  arXiv:0809.2409 [hep-ph].
  I.~Cholis, L.~Goodenough, D.~Hooper, M.~Simet and N.~Weiner,
  arXiv:0809.1683 [hep-ph].
  D.~Hooper and K.~Zurek,
  arXiv:0902.0593 [hep-ph].

\bibitem{Baltz:1998xv}
  E.~A.~Baltz and J.~Edsj\"{o},
  Phys.\ Rev.\  D {\bf 59}, 023511 (1999)
  [arXiv:astro-ph/9808243].

\bibitem{Kamionkowski:1990ty}
  M.~Kamionkowski and M.~S.~Turner,
  Phys.\ Rev.\  D {\bf 43}, 1774 (1991).
  M.~S.~Turner and F.~Wilczek,
  Phys.\ Rev.\  D {\bf 42}, 1001 (1990).


  \bibitem{Moskalenko:1999sb}
  I.~V.~Moskalenko and A.~W.~Strong,
  Phys.\ Rev.\  D {\bf 60}, 063003 (1999)
  [arXiv:astro-ph/9905283].

  \bibitem{GALPROP}
    http://galprop.stanford.edu/web\_galprop/
galprop\_home.html.
    
  \bibitem{Us DM}
    R.~C.~Cotta, J.~S.~Gainer, J.~L.~Hewett, and T.~G.~Rizzo.  In preparation.

\end{thebibliography}
\end{document}